\documentclass[aps,prl,twocolumn,showpacs,floatfix]{revtex4}
\usepackage{graphicx}
\usepackage{amsmath}
\usepackage{amssymb}
\usepackage[utf8]{inputenc}

\def\vec #1{{\bf #1}}
\newcommand{\ie}{\emph{i.}$\,$\emph{e.}}

\begin{document}
\title{Packing of elastic wires in spherical cavities}

\author{ N. Stoop$^1$, J. Najafi$^2$, F. K. Wittel$^1$, M. Habibi$^{2}$, and H. J. Herrmann$^{1,3}$} 
\affiliation{$^1$ Computational Physics for Engineering Materials, ETH Zurich, Schafmattstr. 6, HIF, CH-8093 Zurich, Switzerland\\
$^2$ Department of Physics, Institute for Advanced Studies in Basic Sciences (IASBS), Zanjan 45137-66731, Iran
\\
$^3$ Departamento de F\'isica, Universidade Federal do Cear\`a, Campus do Pici, 60451-970 Fortaleza, Cear\'a, Brazil
} 
\date{April 13, 2011, accepted: April 26, 2011}
   
\begin{abstract}
We investigate the morphologies and maximum packing density of thin wires packed into spherical cavities. Using simulations and experiments, we find that ordered as well as disordered structures emerge, depending on the amount of internal torsion. We find that the highest packing densities are achieved in  low torsion packings for large systems, but in high torsion packings for small systems. An analysis of both situations is given in terms of energetics and comparison is made to analytical models of DNA packing in viral capsids.\end{abstract}
 

\maketitle

Thin objects are ubiquitous in nature and technology. Driven by forces and external constraints, they can undergo surprisingly complex spatial rearrangements, observed, for example, in the folding of insect wings in cocoons, crumpled wires and paper, or growing tissue \cite{Donato:PhysRevE:2002,Stoop:wire,BenAmar:PRoySocLondAMat:1997,Witten:RevModPhys:2007,Blair:PhysRevLett:2005,Dervaux:PhysRevLett:2008,Stoop:econe,Sharon:Nature:2002}. 
%
The packing of long, slender objects in cavities emerges in many situations in biology and mechanics. It occurs in paper jams, when chromatin is stored in the cell nucleus, DNA is injected into viral capsids \cite{Katzav:ProcNatlAcadSci:2006,AliDNA,Purohit:ProcNatlAcadSci:2003}, or when endovascular coils are formed in aneurysm surgery \cite{lanzer2007mastering}. Often, such systems are geared towards high packing densities. For instance, the amount of DNA packed in a capsid limits the genetic information the virus can spread, and in endovascular surgery, high densities improve the long-term stability of the treatment \cite{Tamatani:AmJNeuroradiol:2002}. 
\par
In this Letter, we consider the packing of a thin wire in a rigid spherical container. First, the emergence of ordered as well as disordered packings depending on the amount of internal torsion is shown. We show how the maximum packing density depends on morphology and system size. We find that ordered packings provide higher maximum packing densities at large system sizes, while disordered ones offer higher maximum densities at small system sizes.
\par
We consider two setups in simulations and experiments: In the first, straight wires are used that can axially rotate at the injection point, thereby having minimal torsion during the entire packing process. We call this the \textit{low torsion setup}. In the second setup, we aim at packings that build up internal torsion.
To hinder the release of torsion via the free end, we inject precurved
wires \cite{Landau:1991}. While this modification changes the bending energy, it
surprisingly had no influence on morphology and packing density,
as torsion is released via the cavity opening. We thus also suppress
axial rotation at the opening to arrive at a \textit{high torsion setup}.

\begin{figure*}[t!!!!]
  \begin{center}
   \includegraphics[width=15.5cm]{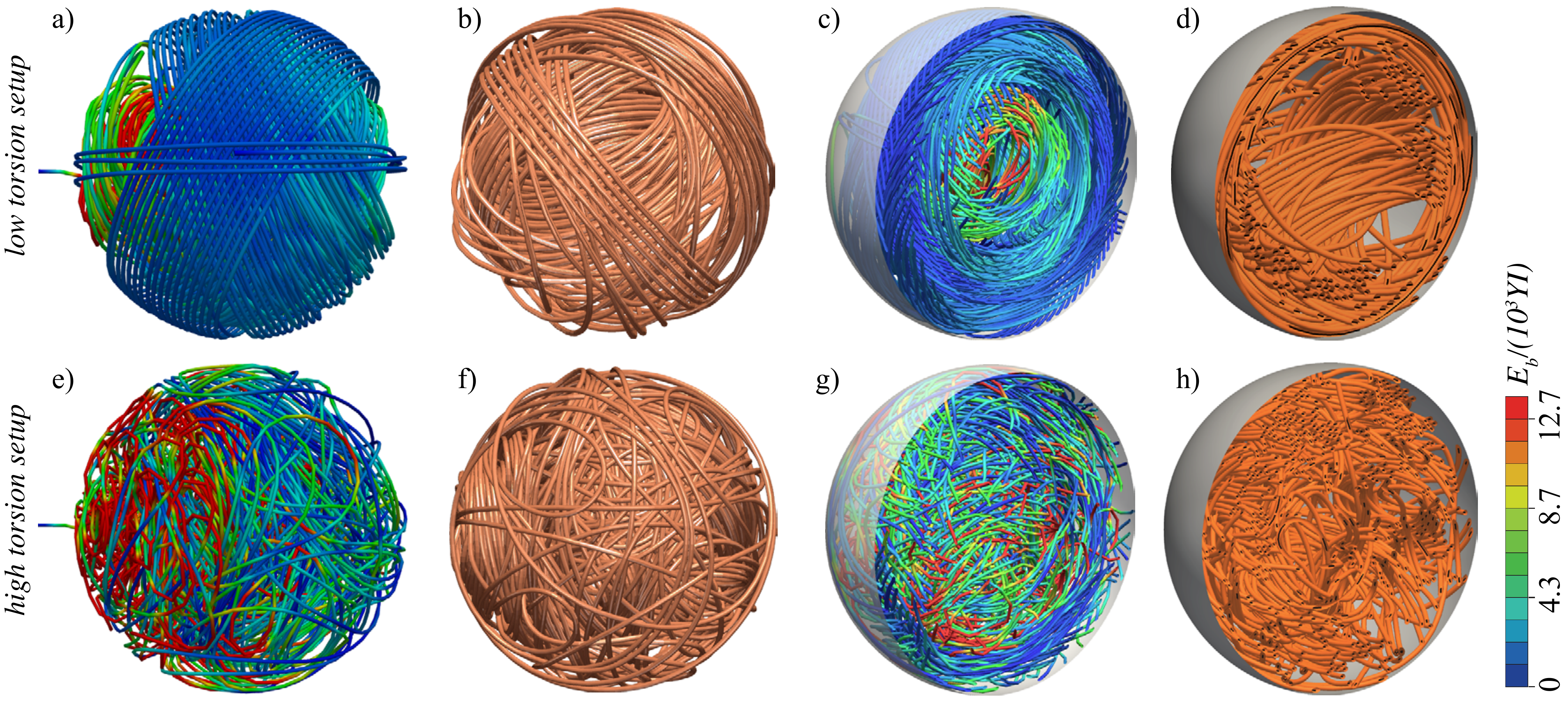}
    \caption{\label{fig:1} \small Morphologies of wires packed into spherical cavities. Top row: The \textit{low torsion setup} results in an ordered morphology, characterized by ring-like coiling (a). A cut through the packing (c) reveils the shell-like inner structure. X-ray tomography scans from experimental realizations are shown in (b, d). Bottom row: The \textit{high torsion setup} produces disordered structures (e, g), with corresponding experiments in (f,h). Color represents the bending energy $E_b$. System parameters are: $\phi=0.23$, $a/R=0.02$ (simulations); $\phi=0.14$, $a/R=0.025$ (low torsion, experiment); $\phi=0.18$, $a/R=0.017$ (high torsion, experiment)%
    }
  \end{center}
\end{figure*}
%

The experiment consists of a transparent, hollow rigid sphere of inner radius $R$ with a small hole to which a nozzle is attached. We insert a Nylon or Silicon wire of constant cross section radius $a$ by two counter-rotating rollers, which allow for a controlled insertion speed and large forces. The Young's modulus $Y$ of the wires was determined from axial extension tests and deflection measurements \footnote{We measured the downward deflection $y$ of the free end of horizontally clamped pieces with different lengths $L$. $Y$ follows by comparison to the analytical result $y=8\lambda g L^{4}/(\pi Y a^{4})$ of the linear rod theory \cite{Landau:1991}}. For the high torsion setup, we use wires of constant intrinsic radius of curvature $R_i \approx 2 R$. In the low torsion setup, we straightened the wires and allowed axial rotation between the nozzle and the sphere. The packing process is recorded by a camera, and the insertion force is measured continuously by a load cell. The process stops when the force becomes so high that no further insertion is possible. Tomography images are taken from the final packing to reconstruct the internal structure.
%
\par
For the numerical model, we describe the wire by its centreline $\gamma$, and an orthonormal director field $\vec{d}_i$, $i=1,2,3$ that specifies the orientation of the cross sections along $\gamma$ \cite{Cosserat:1909}. We parametrize $\gamma$ by $s$, $0 \leq s \leq L$, with $L$ the length of the wire. The position of a point on the centreline is denoted as $\vec{r}(s)$. The director field is oriented such that $\vec{d}_3$ coincides with the centreline tangent, $\vec{r}'(s)$, where $'$ is short-hand for $\partial_s$. It follows from the theory of space-curves that a vector $\vec{k}$ exists such that $\vec{d}'_i = \vec{k} \times \vec{d}_i$. $\vec{k}$ is known as the \textit{Darboux} vector and we denote its components in the director frame by $k_i$. To distinguish between the deformed and undeformed wire configuration, we use the superscript $^0$, \ie{} the reference configuration is described by the smooth curve $\gamma^0$, directors $\vec{d}_i^0$ and Darboux vector $\vec{k}^0$. If the wire is thin compared to its curvature, the internal elastic energy is given by 
\begin{multline}\label{rod_energy}
E_{el} = \frac{1}{2} \int_0^L ds \left\{ YA \left( r_3' - r'^{0}_3 \right)^2 \right.\\
 + YI\left[ (k_1 - k_1^0)^2 + (k_2 - k_2^0)^2 \right] 
\left. + G (k_3-k_3^0)^2 \right\}\;.
\end{multline}

The first term accounts for axial stretching/compression, with the cross section area  $A=\pi a^2$. The second term is the bending energy $E_b$, with $I=\pi a^4/4$ the 2nd moment of inertia. The components $k_1^0$ and $k_2^0$ describe the intrinsic curvature of the reference configuration. The last term accounts for torsion, with $G=\frac{Y}{2(1+\nu)}$ being the torsional shear modulus, and $\nu$ the Poisson ratio. For naturally straight rods, the bending energy simplifies to the well-known expression $\frac{1}{2} \int_0^L YI \kappa^2 ds$, where $\kappa$ is the curvature of the centreline. 
\par
For an efficient simulation including self-contact, we discretize the wire into $N$ mass-points $\vec{r}_i$, connected by straight edges $\vec{e}_i$. Tensile energy is modeled by linear springs connecting mass-points. For bending and torsion, we use the quaternion group to represent the rotation of the director frame of each edge, with resulting moments and forces given by the gradient of the discretized form of Eq.~\ref{rod_energy} \cite{Spillmann:ProceedingsOfThe2007AcmSiggraphEurographicsSymposium:2007}. Contact of the wire with itself or with the cavity is modeled by a linear repelling force law. We use the same stick-slip friction model as in Ref.~\cite{Stoop:wire} with friction coefficients $\mu_{s}=0.2$ for static and $\mu_d=0.18$ for dynamic Coulomb friction - a choice which yields best agreement with experiments. No friction is applied to wire segments in the injection nozzle for both setups. To prohibit axial rotation in the high torsion setup, any resulting axial moment in the nozzle is canceled by an equal but opposing one at every time step.
Forces and moments are integrated in time using a standard predictor-corrector method of 6th order, with small viscous damping added for equilibration and numerical stability. 
We fixed the wire radius to $a=1$ and chose 7 different sphere radii $R=(4$, $5$, $10$, $13$, $20$, $40$, $50)$. The Young's modulus was set to $Y=5$ and $k_y=0.001$ for the high torsion setup, corresponding to an intrinsic radius of curvature of $R_i=100$. The insertion speed was sufficiently small to ensure being in the quasi-static regime, and the simulation was stopped when the measured injection force reached a fixed threshold. 

\begin{figure}[tb]
  \begin{center}
   \includegraphics[scale=0.37]{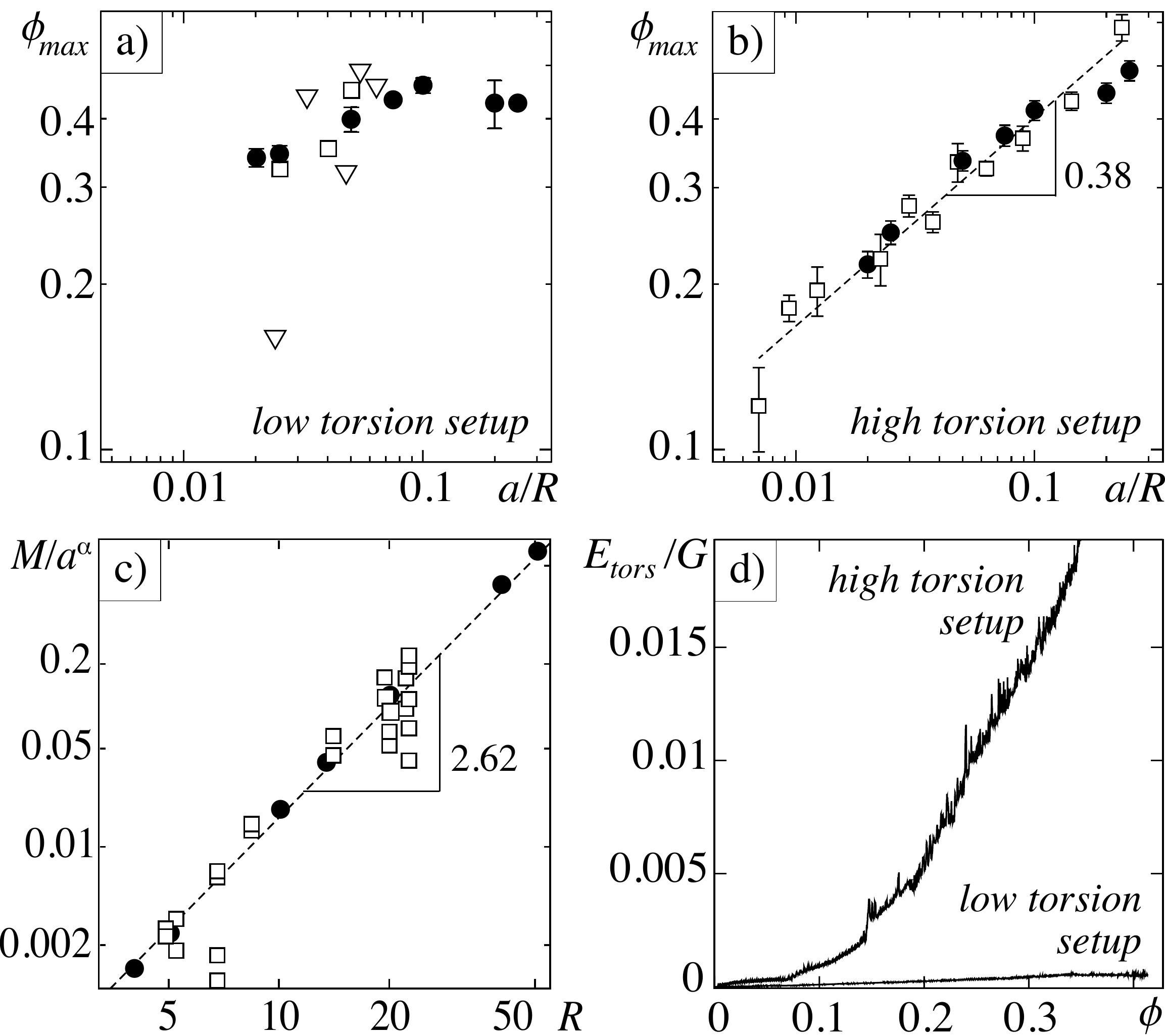} 
    \caption{\label{fig:2} \small Top: Maximum packing density $\phi_{max}$ as a function of the effective system size $a/R$ for the low torsion (a) and the high torsion setup (b). Experimental data are shown with empty, numerical data with filled symbols. DNA packing densities ($\triangledown$) are added in the low torsion setup for reference \cite{Purohit:BiophysJ:2005}. 
Bottom: Length-radius scaling in the high torsion setup (c), and a comparison between the total accumulated internal torsional energy $E_{tors}$ as function of $\phi$ (d).}
  \end{center}
\end{figure}
%
\par
The packing process starts with one end of the wire being inserted into the cavity. To break symmetry of the straight wires in simulations, small random displacements are initially imposed on all nodes. When the wire contacts the cavity walls for the first time, a loop forms along a cavity wall. The orientation of this loop is random for naturally straight wires depending on initial conditions, whereas for curved wires it is in the preferred, curved direction. As more wire is inserted, distinct morphologies emerge for the two setups:
\par
\textit{Low torsion setup:} Here, the wire continues to form loops which align in parallel with each other. This process leads to ring-like structures with decreasing coiling radius. Whenever the coiling radius becomes too small, a new coil is started at a different orientation (Fig.~\ref{fig:1}, a,b). The result is an \textit{ordered packing} in layers from outside inwards (Fig.~\ref{fig:1}, c,d) at earlier stages of the process, similar to the coaxial coiling model for DNA inside spherical capsids \cite{JStructBiol160241}. At higher densities, coils prefer
alignment perpendicular to the feeding axis, sharing similarities with the inverse spool DNA model \cite{Purohit:ProcNatlAcadSci:2003}. 
\par
\textit{High torsion setup:} In this setup, torsion can only be minimized on the expense of bending deformations, leading to frequent reorientations of loops. In certain cases, torsion becomes so large that figure-eight patterns appear, or loops form with smaller radius than imposed by geometric conditions. As a consequence, the packing is \textit{disordered} (Fig.~\ref{fig:1}, e,f) and fills the cavity homogeneously (Fig.~\ref{fig:1}, g,h) with a large amount of accumulated torsion (Fig.~\ref{fig:2}, d).
\par
We turn to the study of the maximum packing density $\phi_{max}$ and its dependence on the effective system size $a/R$. $\phi_{max}$ follows from the inserted wire length $L$ as $\phi_{max}=L \pi a^2 / (4/3 \pi R^3)$ and is shown in Fig.~\ref{fig:2} (a,b) as function of $a/R$ for both setups. 
The system sizes in experiments were in the range $0.0069<a/R<0.23$, with values of $R[mm]=(6.8$, $14.0$, $19.9$, $22.9$, $28.5$, $30.3$, $38.5$, $48.5)$ and $a[mm]=(0.16$, $0.25$, $0.35$, $0.4$, $0.5$, $1.5$, $1.93$, $2.0$, $3.25)$. In simulations $a/R$ was within $0.02$ and $0.25$. Since data points in each setup collapse on a common curve, we conclude that $\phi_{max}$ in either setup only depends on $a/R$ and not on $a$ and $R$ individually. Furthermore, no dependence on Young's modulus $Y$ and on the amount of intrinsic curvature was found within the tested range $R\leq R_i \leq 5R$.
\par
In comparison, the low torsion setup (Fig.~\ref{fig:2}, a) yields higher maximum packing densities for large systems (small $a/R$), comparable with experimental data from DNA coiling (triangles). As $a/R$ is increased, $\phi_{max}$ increases slower than $\phi_{max}$ for the high torsion setup. Consequently, around $a/R=0.2$, the high torsion setup starts to provide larger $\phi_{max}$, with packing fractions up to $58\%$. This could be due to the fact that torsion supports the bending of the wire and thus reduces the required buckling force, similar to the formation of a DNA plectoneme \cite{clauvelin2008mechanical}.
\par
The strong dependence of $\phi_{max}$ on $a/R$ in the high torsion setup is captured well by the power-law $\phi_{max}\sim (a/R)^\alpha$, with $\alpha=0.38\pm 0.04$ (Fig.~\ref{fig:2}, b). We can further conclude that if $\phi_{max}$ scales as a power-law in $a/R$, then the packed wire mass scales as $M \sim a^2 L \sim R^3 (a/R)^\alpha$, \ie{} $M/a^\alpha \sim R^{3-\alpha}$ (Fig.~\ref{fig:2}, c). The exponent $3-\alpha=2.62 \pm 0.04$ is close to the experimental value of $2.75$ found by Gomes et al \cite{aguiar1991geometrical} for the forced crumpling of wires in three dimensions. 
\begin{figure}[tb]
  \begin{center}
   \includegraphics[scale=1.0]{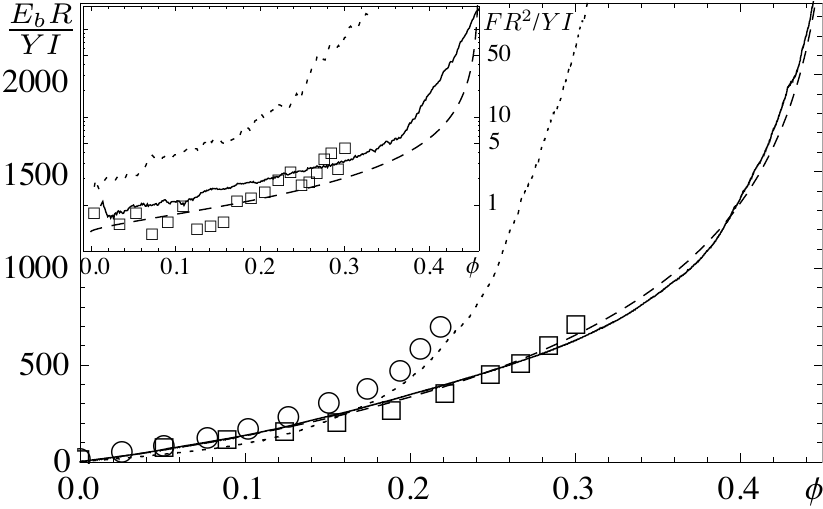}
   \begin{small}
    \caption{\label{fig:3} Rescaled bending energy and injection force (inset) of the wire for the low torsion setup (simulations: continuous line, experiments: {\scriptsize $\Box$}) and the high torsion setup (simulations: dotted line, experiments: {\scriptsize$\bigcirc$}), with $a/R=1/40$. The dashed line is the
       prediction of Ref. \cite{Purohit:ProcNatlAcadSci:2003}. Simulation data are averaged over 9 runs.
         }
         \end{small}
  \end{center}
\end{figure}
\par
The dominant energy in both setups is the bending energy $E_b$. To compare results of different cavity radii,  Fig.~\ref{fig:3} shows the dimensionless quantity $E_b R /YI$, which yields good agreement between simulations and experiments \footnote{Experimentally, an approximation of $E_b$ follows from numerical integration of the measured injection force over $L$.}. We compare these findings to the packaging of DNA in viral capsids. While this process is also accompanied by, e.g., entropic and electronic contributions \cite{PNAS2009}, comparison can be made to the bending-dominated DNA model of Purohit et al. \cite{Purohit:ProcNatlAcadSci:2003}. It predicts 

\begin{equation}\label{purohit_model}
E_b \sim -\frac{R Y I}{d_s^2} \left[ \sqrt{k}\phi^{1/3} + \log\left( \frac{1 - \sqrt{k} \phi^{1/3}}{\sqrt{1-k \phi^{2/3}}}\right)   \right]\;,
\end{equation}
in which $k = [3 d_s^4/ ( 4\pi^2 a^4)]^{1/3}$. The only remaining parameter is the average segment distance $d_s$, which we determined from cross sections of the simulations as $d_s=2.81 a$. The prediction agrees well with our data, c.f. Fig.~\ref{fig:3}, dashed line. The dimensionless insertion force $FR^2/YI$ obtained analytically from Eq.~\ref{purohit_model} is, however, somewhat smaller than the measurements (Fig.~\ref{fig:3}, inset), due to the fact that we include contact friction.
\begin{figure}[t]
  \begin{center}
   \includegraphics[scale=1.03]{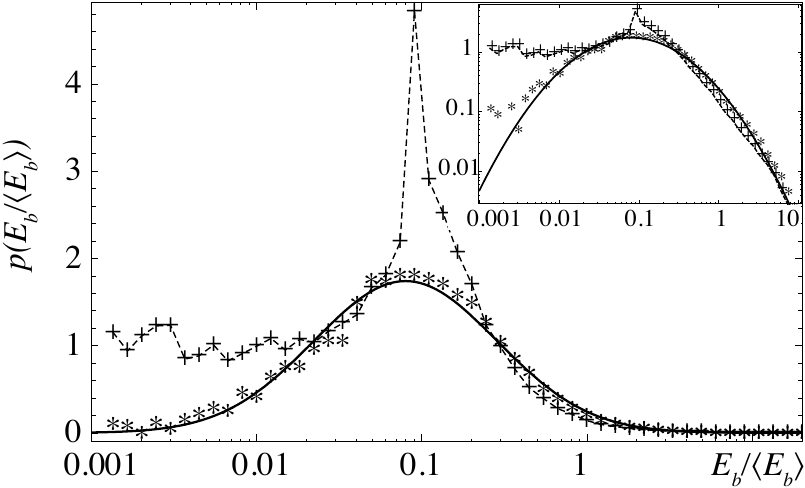}
    \caption{\label{fig:4} \small Bending energy distribution of the packed wire in log-linear scale and log-log scale (inset) for $a/R=1/40$. Data for the high torsion setup ($*$) can be approximated by a log-normal distribution with parameters $\mu=-0.903$ and $\sigma=1.275$ (continuous line). For the low torsion setup ($+$, dotted line is shown as a guide to the eye) small bending energies dominate due to the ordered coiling.}
  \end{center}
\end{figure}
\par
The distribution of the bending energy measured at $\phi_{max}$ gives further insights into the statistical properties of the morphologies (see Fig.~\ref{fig:4}). The low torsion setup ($+$) features a strong concentration of energies around $E_b/\langle E_b \rangle = 0.1$ due to the presence of ordered coils with large radii. Subsequent coiling rings are squeezed into existing ones. They take on a shape reminding of stadium racepaths, with almost straight parts that lead to a high probability for small $E_b$.
In contrast, the high torsion setup ($*$) yields a distribution fitted well by a log-normal function of the type
\begin{equation}
p(x = E_b/\langle E_b \rangle ) = \frac{1}{\sigma x \sqrt{2 \pi}} \exp\left[-\frac{(\ln(x) - \mu)^2}{2 \sigma^2}\right] \; ,
\end{equation}
with $\mu = -0.903$ and $\sigma = 1.275$. Log-normal distributions are usually associated with hierarchical events and are found in similar systems of densely packed objects, for instance in the ridge-length distribution of crumpled paper or two-dimensional crumpled wires \cite{Blair:PhysRevLett:2005, sultan2006statistics, Witten:RevModPhys:2007}.
%
\par
To summarize, we investigated the packing of elastic wires into spherical cavities using a high torsion and a low torsion setup. Low torsion led to ordered packings, while high torsion led to disordered structures. We found highest packing densities in the first case for large cavities and in the second case for small cavities, with a cross-over as the system size is varied. Our work elucidates the importance of torsion in the dense packing regime, and provides novel insights into the role of the system size with relevance for, e.g., the surgical treatment of aneurysms. The presented results are largely independent on Young's modulus, the amount of intrinsic curvature within the tested range, and friction. The role of material nonlinearities, however, remains an important open question in view of such applications.

\begin{acknowledgments}

This work was supported by Grant No. TH-0607-3 of ETH Zurich, FUNCAP, and grant No. G2010IASBS103 of the Institute for Advanced Studies in Basic Sciences (IASBS) Research Council. The authors would like to thank R.L. Stoop, S. Kusuma and EMPA for their help in the analysis part of this work.
\end{acknowledgments}


\end{document}